\documentclass{article}
\usepackage{spconf,amsmath,graphicx,hyperref,booktabs,multirow}

\title{UCONV-CONFORMER: HIGH REDUCTION OF INPUT SEQUENCE LENGTH\\ FOR END-TO-END SPEECH RECOGNITION}
\name{Andrei Andrusenko$^{1}$, Rauf Nasretdinov $^{2}$, Aleksei Romanenko $^{1,2}$}
\address{
  $^1$ITMO University, St. Petersburg, Russia\\
  $^2$STC-innovations Ltd, St. Petersburg, Russia\\
  \texttt{\small{andrusenkoau@itmo.ru, \{nasretdinov, romanenko\}@speechpro.com}}}

\begin{document}
%
\maketitle
\begin{abstract}
Optimization of modern ASR architectures is among the highest priority tasks since it saves many computational resources for model training and inference. The work proposes a new Uconv-Conformer architecture \footnote{\url{https://github.com/andrusenkoau/espnet/blob/master_uconv/espnet2/asr/encoder/uconv_conformer_encoder.py}} based on the standard Conformer model. It consistently reduces the input sequence length by 16 times, which results in speeding up the work of the intermediate layers.
To solve the convergence issue connected with such a significant reduction of the time dimension, we use upsampling blocks like in the U-Net architecture to ensure the correct CTC loss calculation and stabilize network training. The Uconv-Conformer architecture appears to be not only faster in terms of training and inference speed but also shows better WER compared to the baseline Conformer. Our best Uconv-Conformer model shows 47.8\% and 23.5\%  inference acceleration on the CPU and GPU, respectively. Relative WER reduction is 7.3\% and 9.2\% on LibriSpeech test\_clean and test\_other respectively.
\end{abstract}
\section{Introduction}
\label{sec:intro}

The active development of the end-to-end approaches in ASR has led to many robust neural network architectures based on the convolutional neural networks \cite{jasper, quartznet} and the attention mechanism \cite{attention}. A Conformer model \cite{conformer} developed by combining these approaches is currently a state-of-the-art solution for many ASR systems based on both end-to-end \cite{conformer_sota} and hybrid \cite{hybrid-conformer-sat} approaches. However, this model has several drawbacks related to the complexity of training and low inference speed. It mainly occurs due to the quadratic complexity of self-attention with respect to the length of the processed sequence \cite{efficient_transformers}.
As a result, a successful training of the Conformer model and its inference in production take much time and require a lot of computational resources.

Recently, various approaches have been proposed to make the self-attention mechanism more efficient \cite{Longformer, bigbird}. 
The work \cite{branchformer} uses the Fastformer \cite{wu2021fastformer} model with linear complexity, which is based on additive attention. In \cite{katharopoulos_et_al_2020} authors proposed an approach based on kernelization.

Another way to speed up the Conformer is to reduce the input sequence length. The vanilla implementation of the Conformer model has an initial convolution block that reduces the input sequence length by 4 times; we further denote the overall reduction of an input sequence length by N times as a \textit{reduction level xN}. In Conv-Transformer \cite{conv_transformer} and Efficient Conformer \cite{efficient_conformer} the authors decrease the input sequence length by 8 times placing intermediate downsampling blocks after several Conformer layers. This approach significantly reduces the number of calculations for the rest of the model without considerable loss in recognition accuracy.
However, this option works only for subword output units since the reduction level x8 is too high for graphemes. It may lead to a CTC loss \cite{ctc} calculation error because the number of output embeddings may become less than the number of targets in the training transcription. The same problem arises for subwords when the input sequence length is reduced by 16 times. In this case, the effect of losing some latent acoustic information is added due to the rejection of even more intermediate signal representations than it is at reduction level x8. All these factors cause the problem of model convergence.

Inspired by the U-Net architecture 
\cite{unet}, we propose the Uconv-Conformer model -- an architecture for sequentially reducing the input sequence length by 16 times, followed by restoring the reduction level to x8 for the final layers of the Conformer model. To restore a latent acoustic information discarded during progressive downsampling, we use skip connections between blocks with the same reduction level, similar to the U-Net approach. This method allows us to reduce the number of processed embeddings for intermediate Conformer layers while maintaining the final reduction level acceptable for the CTC loss calculation using subwords.

\begin{figure*}[t]
  \centering
  \includegraphics[scale=0.19]{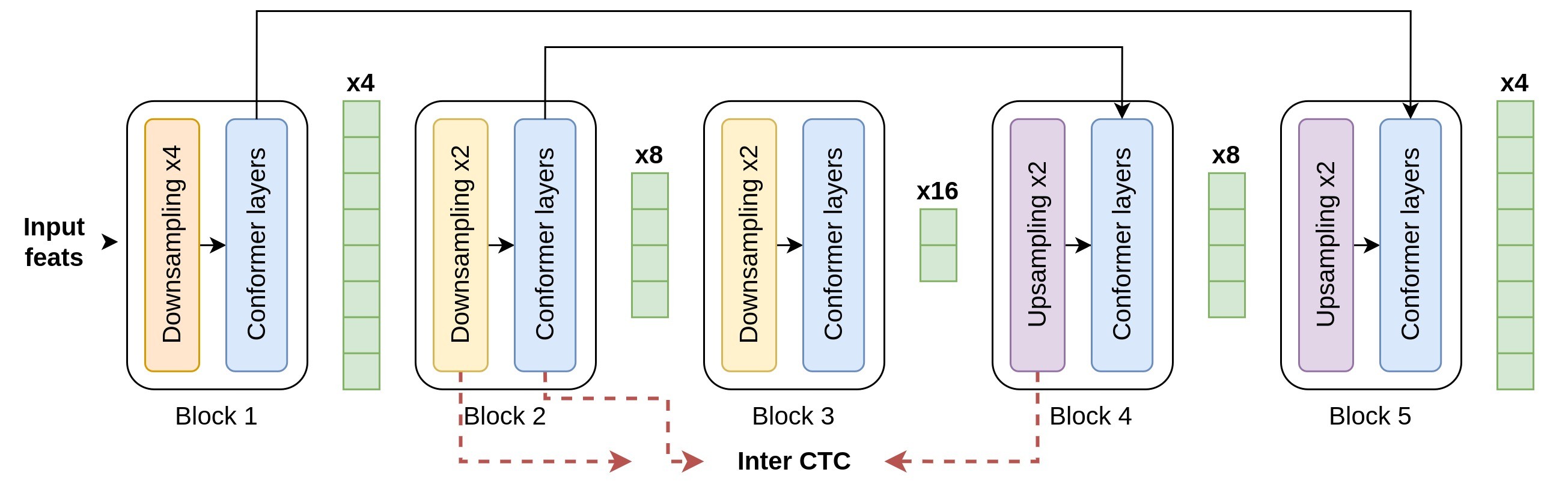}
  \caption{\textit{The Uconv-Conformer architecture.}}
  \label{fig:uconv}
\end{figure*}

In parallel with the active testing of the final prototype of our Uconv-Conformer model, a Squeezeformer \cite{kim2022squeezeformer} paper has been released. The key idea of this work is to use a temporal U-net structure similar to our approach. The main difference is that the Squeezeformer uses the intermediate reduction level x8 and the reduction level x4 for the final layers. Our approach is initially designed for higher reduction levels.

For additional comparison we implemented the Conv-Conformer architecture similar to the Conv-Transformer \cite{conv_transformer} with some simplifications. We also investigated the use of the intermediate CTC loss \cite{interctc} for individual blocks of the Uconv-Conformer architecture. According to our results, it makes the training process more stable and improves the recognition accuracy for the low-resource data scenario.

As a measure of the model speed, we provide an extended comparison of the studied architectures in terms of inference speed on CPU and GPU.

\section{Methods}
\label{sec:format}

\subsection{Uconv-Conformer architecture}

The primary motivation behind the Uconv-Conformer model (Figure~\ref{fig:uconv}) is to reduce the length of embedding sequence processed by the intermediate part of the Conformer model. In the works \cite{conv_transformer} and \cite{efficient_conformer}, it is done by the sequential downsampling of the embedding sequence length by 8 times. According to our experience, reduction level x8 is more effective for subwords in English. This reduction level is too high for graphemes (the average pronouncing duration of a grapheme is less than a subword) and my lead to errors in CTC loss calculation and degradation of recognition accuracy.

It is possible to further increase the reduction level inside the network up to x16 (the largest reduction level in the model is further called as a \textit{reduction depth}, and the reduction level for the final layer is called as a \textit{final reduction}), but this leads to the same problems as the reduction level x8 for graphemes. To avoid this, we suggest using an upsample back to reduction level x8 for the final layers of the model in the case of subwords, that is similar to the U-Net approach. An additional upsampling block can be added for graphemes targets to obtain the final reduction level x4. This can be useful for languages that work better with phonemes or graphemes \cite{openstt_ru}.

However, adding only upsampling blocks is not enough, since some latent acoustic information is lost on the high reduction levels. To recover this loss we apply skip connections between model blocks with the same reduction level (the long dark arrows in Figure~\ref{fig:uconv}). Without using skip connections the recognition accuracy degrades noticeably.

As intermediate downsampling blocks (Downsampling x2), we use a sequence of three conv1d layers like in \cite{conv_transformer}. The first conv1d layer with filter size 3 and stride 1 converts the input dimension to an intermediate one (512 in our work). The second conv1d layer with filter size 3 and stride 2 reduces the input sequence length by 2 times. The final conv1d layer acts as the fully-connected layer and converts the result back to the attention dimension of the Conformer block. For the upsampling, we use the standard upsample module from PyTorch with nearest neighbor algorithm and scale factor 2.

\subsection{Intermediate CTC loss}

The proposed Uconv-Conformer architecture consists of several blocks operating at different reduction levels. To improve the training convergence of such a model, we analyzed the influence of the intermediate CTC loss \cite{interctc} calculated at the output of such separate blocks (the red dashed arrows in Figure~\ref{fig:uconv}). It is necessary to bring all intermediate outputs to the same reduction level (x8 in our case) to calculate CTC loss under the same conditions. Otherwise, the network training stops converging. 
To calculate the final loss function, we use an expression like in \cite{nozaki21_interspeech} with lambda 0.5.

\section{Experimental setup}
\label{sec:pagestyle}

\subsection{Baseline model}

As a starting point, we use the standard 12-layer Conformer encoder from the ESPnet librispeech recipe, which has 83.4M parameters and initial downsampling block x4\footnote{\url{https://github.com/espnet/espnet/blob/master/egs2/librispeech/asr1/conf/train_asr_conformer.yaml}}.
We refer to this model as Conformer-L. According to \cite{hybrid-conformer-sat} and our experience with Conformers, such a large model size may be excessive for training on a small amount of data. Therefore, we reduced the number of model parameters to 21.8M.
Now the number of conv2d convolution filters in the initial Downsampling x4 block is 64. The attention layer size is 280 with eight heads. The feed forward layer size is 1024. The conformer convolution filter size is 5.
Thus we obtained a significant acceleration of the training and testing process without a critical loss in recognition accuracy. This model is referred to as the Conformer-S and used as the main baseline.

\subsection{Conv-Conformer model}

For additional comparison with our approach, we implemented the Conv-Conformer model similar to the Conv-Transformer \cite{conv_transformer}. The main difference is merging two time reduction steps into a single initial Downsampling x4. This model can be described as the first two blocks in Figure~1. We explored two options for the distribution of 12 layers between reduction level x4 and reduction level x8: 2-10 and 4-8.
The more layers there are at reduction level x8, the faster the model operates. The size and number of Conformer layers are the same as in the baseline Conformer-S. The only difference is the presence of an additional Downsampling x2 block, consisting of three consecutive conv1d layers with the same size as in the Uconv-Conformer.

\subsection{Uconv-Conformer model}

We explored different reduction depths and final reduction levels for the proposed Uconv-Conformer model. The recently published Squeezeformer model can be compared to our Uconv\_D8-F4 model in terms of its time reduction policy. The notation D8-F4 denotes that the reduction depth (D) is x8, and the final reduction (F) is x4. We also considered a deeper downsampling version -- Uconv\_D16-F4 with the same final reduction level x4. Models Uconv\_D16-F8\_v1 and Uconv\_D16-F8\_v2 have reduction depth x16 and final reduction x8. However, they have a different number of layers at each of the reduction levels. By varying the number of layers between different reduction levels it is possible to increase the model speed at the expense of its recognition accuracy and vice versa. The size and number of Conformer layers are the same as in Conformer-S. The number of parameters differs due to two additional Downsampling x2 blocks, each having about 1.4M parameters. For additional comparison with large Conformer-L we use Uconv-L\_D16-F8\_v1, which has the same size of the Conformer layers except for the additional Downsampling x2 blocks.
Table~\ref{tab:results_100h} describes the parameters, time reduction policy, and layers distribution at different reduction levels.

\subsection{Training details}

To evaluate different scenarios with respect to the amount of available training data, we use the standard LibriSpeech \cite{librispeech} division into 100, 460, and 960 hours of training data. We use 80-dimensional Mel-scale log filter banks as acoustic features, calculated with 25ms window and 10ms shift. SpecAugment \cite{specaugment} is applied during the training process to prevent model overfitting. As acoustic targets, we use 256 BPE \cite{bpe} unints obtained with SentencePiece tokenizer \cite{sentencepiece}.

All models were trained for 100 epochs with Adam optimizer and weight decay 1e-6. Warmup with a peak LR value of 2e-3 at the 15th training epoch was used as a scheduler. We used the numel batch type with fixed batch bins 2M (sorting audio samples by length to reduce calculations on padded parts) and gradient accumulation is equal to 6.

CTC loss was used as the objective function. Decoding was performed using the beam search algorithm with beam 20 without using any external language model. All experiments were carried out in the ESPnet toolkit \cite{watanabe2018espnet}.

\subsection{Measuring model speed performance}

We provide the inference speed of the resulting model on the GPU and the CPU since many scenarios for using ASR systems still involve using the CPU. To do this, we measure the processing time of a 30-second test sample launched in one thread on the Intel Core i7-8700 and the single RTX 2080 Ti.

\section{Results}
\label{sec:typestyle}

\begin{table*}[ht]
  \centering
  \small
  \caption{\textit{Results (WER, \%) for training on LibriSpeech 100h. Layers distribution shows the number of Conformer layers at each reduction level. Inference speed on CPU/GPU presented in comparison with Conformer-S model. A negative value indicates an improvement, and a positive value -- a degradation by X percent.}}
  \medskip
  \label{tab:results_100h}
  \begin{tabular}{l || c | c | c || c | c || c c c c}
    \toprule
    \multirow{2}*{\textbf{Model}} & \textbf{Params} & \textbf{Reduction} & \textbf{Layers} & \multicolumn{2}{c||}{\textbf{Inference speed \%}} & \textbf{dev} & \textbf{dev} & \textbf{test} & \textbf{test} \\
    & \textbf{M} & \textbf{policy} & \textbf{distrib.} & \textbf{CPU} & \textbf{GPU} & \textbf{clean} & \textbf{other} & \textbf{clean} & \textbf{other}\\
    \midrule
    Conformer-S (\textbf{baseline})& 21.8 & x4 & 12 & - & - & 10.8 & 28.8 & 11.3 & 28.9 \\
    Conv-Conformer\_v1 & 23.2 & x4-x8 & 2-10 & -48.4 & -27.3 & 11.4 & 28.5 & 11.6 & 29.3 \\
    Conv-Conformer\_v2 & 23.2 & x4-x8 & 4-8 & -38.5 & -25.0 & 11.2 & 28.0 & 11.5 & 28.4 \\
    Uconv\_D8-F4 & 23.2 & x4-x8-x4 & 2-8-2 & -38.3 & -17.8 & 11.1 & 28.2 & 11.2 & 29.2 \\
    Uconv\_D16-F4 & 24.6 & x4-x8-x16-x8-x4 & 2-2-4-2-2 & -44.2 & -20.7 & 12.4 & 30.2 & 12.6 & 30.8 \\
    Uconv\_D16-F8\_v1 & 24.6 & x4-x8-x16-x8 & 3-3-3-3 & -47.8 & -23.5 & \textbf{10.4} & \textbf{27.1} & \textbf{10.7} & \textbf{27.1} \\
    Uconv\_D16-F8\_v2 & 24.6 & x4-x8-x16-x8 & 2-4-5-1 & -56.3 & -25.2 & 10.9 & 27.4 & 11.0 & 27.9 \\
    Uconv\_D32-F8 & 26.0 & x4-x8-x16-x32-x16-x8 & 2-2-2-2-2-2 & -58.1 & -27.7 & 11.5 & 28.6 & 11.6 & 29.0 \\
    \midrule
    Conformer-L & 83.0 & x4 & 12 & +221.8 & +113.3 & 10.7 & 27.8 & 10.9 & 28.4 \\
    Uconv-L\_D16-F8\_v1 & 87.3 & x4-x8-x16-x8 & 3-3-3-3 & +124.7 & +71.2 & \textbf{10.4} & \textbf{26.7} & \textbf{10.7} & \textbf{27.2} \\
    \bottomrule
  \end{tabular}
\end{table*}

Table~\ref{tab:results_100h} provides the results for the 100h training data scenario. It can be seen that Conformer-S is quite a bit inferior to Conformer-L in recognition accuracy but significantly outperforms it in inference speed. The Conv-Conformer models show comparable WER results to Conformer-S, but have an advantage over it in inference time.

Next, we explored the different reduction depths and final reduction levels for Uconv-Conformer. The results show that all the models are faster than Conformer-S baseline, but returning to the final reduction level x4 deteriorates recognition accuracy. At the same time, the Uconv-Conformer models with final reduction level x8 show substantial performance boost and provide the best WER. Arranging more layers at reduction levels x8 and x16 allows to accelerate the model in terms of CPU inference speed. However, a slight accuracy degradation occurs due to the loss of some acoustic information at a high reduction level. Noteworthy, Uconv\_D16-F8\_v1 consumes about 28\% less GPU memory during training than Conformer-S, which allows to increase the batch size and obtain additional training speed acceleration.


Table~\ref{tab:results_460_960h} presents the experiments with all major architectures for 460 and 960 hours of available training data. Uconv\_D16-F8\_v1 outperforms Conformer-S and Conv-Conformer\_v2 in all the cases. Our model even shows comparable results with Conformer-L despite much fewer model size and higher work speed. Furthermore, large Uconv-L\_D16-F8\_v1 demonstrates that Uconv-Conformer architecture shows a consistent superiority in recognition accuracy over standard Conformer  in case of comparable model size.

\begin{table}[hb]
  \centering
  \small
  \caption{\textit{Results (WER, \%) for small and large models trained on LibriSpeech 460h and 960h.}}
  \medskip
  \label{tab:results_460_960h}
  \begin{tabular}{l | c c c c}
    \toprule
    \multirow{2}*{\textbf{Model}} & \textbf{dev} & \textbf{dev} & \textbf{test} & \textbf{test} \\
    & \textbf{clean} & \textbf{other} & \textbf{clean} & \textbf{other} \\
    \midrule
    \midrule
    \multicolumn{5}{c}{\textbf{LibriSpeech 460h}} \\
    \midrule
    \midrule
    Conformer-S (\textbf{baseline})& 5.1 & 16.9 & 5.5 & 16.8 \\
    Conv-Conformer\_v2 & 5.2 & 16.4 & 5.4 & 16.2 \\
    Uconv\_D16-F8\_v1 & \textbf{4.8} & \textbf{15.7} & \textbf{5.0} & \textbf{15.2} \\
    Uconv\_D16-F8\_v2 & 4.9 & 15.7 & 5.1 & 15.6 \\
    \midrule
    Conformer-L & 4.8 & 15.6 & 5.0 & 15.2 \\
    Uconv-L\_D16-F8\_v1 & \textbf{4.5} & \textbf{15.0} & \textbf{4.7} & \textbf{14.5} \\
    \midrule
    \midrule
    \multicolumn{5}{c}{\textbf{LibriSpeech 960h}} \\
    \midrule
    \midrule
    Conformer-S (\textbf{baseline})& 4.1 & 11.0 & 4.1 & 10.9 \\
    Conv-Conformer\_v2 & 4.0 & 10.6 & 4.1 & 10.6 \\
    Uconv\_D16-F8\_v1 & \textbf{3.7} & \textbf{10.1} & 3.8 & \textbf{9.9} \\
    Uconv\_D16-F8\_v2 & 3.8 & 10.2 & \textbf{3.7} & 10.0 \\
    \midrule
    Conformer-L & 3.6 & 9.9 & 3.7 & 9.7 \\
    Uconv-L\_D16-F8\_v1 & \textbf{3.4} & \textbf{9.1} & \textbf{3.5} & \textbf{9.0} \\
    \bottomrule
  \end{tabular}
\end{table}

\begin{table}[!h]
  \centering
  \small
  \caption{\textit{Relative WER change, \% of Uconv\_D16-F8\_v1 model after using intermediate CTC loss for different sizes of training data set.}}
  \medskip
  \label{tab:results_100h_inter_ctc}
  \begin{tabular}{l | c c c c}
    \toprule
    \multirow{2}*{\textbf{Training data}} & \textbf{dev} & \textbf{dev} & \textbf{test} & \textbf{test} \\
    & \textbf{clean} & \textbf{other} & \textbf{clean} & \textbf{other} \\
    \midrule
    LibriSpeech 100h & \textbf{-6.7} & \textbf{-2.2} & \textbf{-6.5} & \textbf{-1.7} \\
    LibriSpeech 460h & +2.1 & +4.5 & +4.0 & +4.6 \\
    LibriSpeech 960h & +8.1 & +5.9 & +10.5 & +9.1 \\
    \bottomrule
  \end{tabular}
\end{table}

Using an intermediate CTC loss for the Uconv-Conformer model (Table~\ref{tab:results_100h_inter_ctc}) improves the recognition accuracy only for 100h training data scenario. It leads to quality degradation in case of 460h and 960h training data. We can conclude that the intermediate CTC loss is only helpful for the low-resource scenario as an additional regularization. 

\section{Conclusion}
\label{sec:majhead}

This paper proposes the Uconv-Conformer architecture, which reduces the input sequence length by 16 times for intermediate Conformer layers, followed by upsampling to final reduction level x8 aimed at stabilization of convergence and CTC loss calculation.
We also investigated the effect of the intermediate CTC loss, which improves the recognition accuracy for the low-resource scenario.
The results obtained for small models trained on LibriSpeech 960 hours outperform the baseline Conformer with a similar number of parameters on test\_clean and test\_other by 7.3\% and 9.2\% relative WER. 
This method also accelerates inference on CPU and GPU by 47.8\% and 23.5\%, respectively.

\section{ACKNOWLEDGEMENTS}
\label{sec:acknow}
This work was supported by the Analytical Center for the Government of the Russian Federation (IGK 000000D730321\ P5Q0002), agreement No. 70-2021-00141.




\bibliographystyle{IEEEbib}
\bibliography{mybib}

\end{document}